\documentstyle[aps,preprint,eqsecnum,prb]{revtex}
\begin{document}
\draft
\title{Theory of phase-locking in small Josephson junction cells}

\author{M. Basler\thanks{pmb@rz.uni-jena.de},
W. Krech\thanks{owk@rz.uni-jena.de} and
K. Yu. Platov\thanks{okp@rz.uni-jena.de}
}
\address{
\sl Friedrich-Schiller-Universit"t Jena\\
\sl Institut fr Festk"rperphysik\\
\sl 07743 Jena\\
\sl Max-Wien-Platz 1}
\date{February 13, 1995}
\maketitle
\begin{abstract}
Within the RSJ model, we performed a theoretical analysis of
phase-locking in elementary strongly coupled Josephson junction cells.
For this purpose, we developed a sy\-ste\-ma\-tic method
allowing the investigation of
phase-locking in cells with small but non-vanishing loop inductance.
The voltages across the junctions are found to be locked
with very small phase difference for almost all values of external
flux.
However, the general behavior of phase-locking is found to be just
contrary to that according to weak coupling. In case of strong coupling
there is nearly no influence of external magnetic flux on the
phases, but the locking-frequency becomes flux-dependent.
The influence of parameter splitting is considered as well as the
effect of small capacitive shunting of the junctions.
Strongly coupled cells show synchronization even for large
parameter splitting.
Finally, a study of the behavior under external microwave radiation
shows that the frequency locking-range becomes strongly
flux-dependent, whereas the locking frequency itself turns out to be
flux-independent.
\end{abstract}

\section{Introduction}
Josephson junction arrays are considered as
candidates for microwave oscillators with possible applications
in the field of satellite communication systems,
astronomical observations, construction of supercomputer chips and
spectroscopy
\cite{jain1,hansen1,likharev1,benz1,benz2,lindelof1,lukens1,lukens2,konopka1}.
They are potentially well tunable over a
relatively wide frequency range while radiating on a narrow linewidth,
and they can deliver large output power, at least
in comparison with a single element. Linear
arrays of Josephson junctions have been subject to
theoretical investigation for more than twenty years
\cite{jain1,likharev1,lukens2,krech6,krech7,hadley1}.
During last years, there have been some promising experimental
results\cite{wan1,han1,han2}. Since the beginning of the nineties,
there has been a growing interest in 2D arrays
\cite{benz1,benz2,booi1,martens1,stern1,liu1,doderer1,kautz1}.
Up to now, the radiation power of 2D arrays was found to be
much smaller than that of 1D arrays
(0,1 $\mu$W maximum\cite{stern1}, in comparison to 50
$\mu$W in 1D arrays\cite{han2}). This may be a consequence of
technological problems as well as of general properties of 2D arrays.

Because of this fact, there is some renewed interest in the general
mechanisms of synchronization of coupled Josephson junctions. Most of
the early adiabatic investigations
were based on weak (preferably inductive) coupling of the
elements\cite{jain1,likharev2,kuzmin1,krech3,krech5,krech8,volkov1,stephen1},
which is surely fulfilled for
relatively large circuits with total
inductances $\stackrel{>}{\sim} 10$ pH.
However, present day technology allows the preparation of very compact
arrays with circuit dimensions around 1 $\mu$m and smaller, having
inductances smaller than 3pH. In this case, the adiabatic methods
developed earlier fail. On the other hand, neglecting inductances at
all\cite{sohn1} seems to be a too rough approximation.

In this paper we develop a method for dealing with
Josephson junctions coupled via a small inductive shunt
(SQUID-type coupling) in a systematic way.
In Sec. II, we describe the
circuits handled by this method, derive the general equations,
review some results obtained by conventional methods and
discuss their inapplicability to our problem. In Sec. III,
we give a review of our analytical method and present some results for
the simplest case of non-hysteretical,
identical junctions, which are compared with
numerical simulations.
Sec. IV is devoted to non-identical junctions and Sec. V
to the influence of
an additional small junction capacitance.
In Sec. VI we consider the synchronization of strongly
coupled SQUID cells by
external microwave radiation.
A summary as well as some speculations about
possible applications of our results are given in Sec. VII.

Contrary to most theoretical investigations which are mainly
based on computer simulations during the last years,
we concentrate on approximate
analytical results. The advantage of this approach is usually a better
insight into physical mechanisms in combination with a broader range of
applicability concerning the choice of parameters. The disadvantage is
a larger amount of mathematical machinery necessary even in relatively
simple cases. Howsoever, we found it quite valuable to use both
methods and compare the results. Some material described in Sec. III
was published in a short note earlier\cite{basler1}.

\section{A short review on weakly coupled Josephson junctions}
The strong coupling method described in this paper was
developed for the investigation of the
three similar SQUID-like cells shown in Fig. \ref{fig1}, but the
general principles should have a much wider range of applicability.
These three circuits have a bias current $2I_0$, a net loop
inductance $L$ and a parallel biasing scheme in common.
Elaborating the equations of motion within the RSJ model, one obtains
for the case of identical junctions
\begin{mathletters}
\begin{eqnarray}
 \dot{\phi_1}+\sin\phi_1&=&i_0-\l^{-1}(\phi_1-
\phi_2+\varphi),\label{rsj1}\\
 \dot{\phi_2}+\sin\phi_2&=&i_0+\l^{-1}(\phi_1-
\phi_2+\varphi)\label{rsj2}
\end{eqnarray}
\end{mathletters}
where $\phi_1$ and $\phi_2$ are the respective Josephson phases, and
 $i_0$ is the normalized bias current,
\begin{equation}
 i_0=I_0/I_C\quad\mbox{($I_C$: critical current)},
\end{equation}
which is supposed to fulfill the condition $i_0>1$, here.
$l$ marks the normalized loop inductance,
\begin{equation}
 l=2\pi I_CL/\Phi_0,
\end{equation}
and $\varphi$ the normalized external flux,
\begin{equation}
 \varphi=2\pi\Phi/\Phi_0\quad\mbox{($\Phi$: external flux,
$\Phi_0$: flux quantum)}.
\end{equation}
Dots denote differentiation  w.r.t. the scaled time
\begin{equation}
s=(2e/\hbar)I_CR_Nt\quad\mbox{($R_N$: normal resistance)}.
\end{equation}

There exist several investigations of
these systems for weak coupling, i.e.
$l\gg1$\cite{jain1,likharev1,krech8}. In this case, coupling can be
neglected to $0^{th}$ order w.r.t. $l^{-1}$
and both junctions oscillate with the Josephson
phase of an overcritically biased free contact,
\begin{equation}\label{free}
 \phi_{1/2}=\mbox{arctan}\left[\frac{\zeta_0}{i_0+1}
 \tan\Big(\frac{\zeta_0s-\delta_{1/2}}{2}\Big)\right]+\frac{\pi}{2},
\end{equation}
showing an oscillation frequency
\begin{equation}
 \zeta_0=\sqrt{i_0^2-1}
\end{equation}
and constant phases $\delta_1$ and $\delta_2$. To $1^{st}$
order the lowest harmonics of solutions (\ref{free}) are inserted
into the r.h.s. of Eqs. \ref{rsj1},b.
This provides so-called reduced equations for the
mean values of the phases, averaged over short time scales
of the order of $\zeta_0$. Looking for solutions of the
phase-locking type,
\begin{equation}
 \dot{<\delta>}=0,\quad<\delta>=<\delta_1>-<\delta_2>,
\end{equation}
one finds
\begin{equation}
  <\delta>-\frac{1}{i_0\Big(i_0+\zeta_0\Big)}\sin<\delta>
 =\varphi.
\end{equation}
For usual operation regimes with $i_0\approx1.5$ this
results in an approximate linear relation between $\delta$ and $\phi$,
as indicated in Fig. \ref{fig2}. The normalized voltages
\begin{equation}
 v_{1/2}=\frac{V_{1/2}}{I_CR_N}
\end{equation}
are obtained as
\begin{equation}
 v_{1/2}=\frac{\zeta_0^2}{i_0+\cos(\zeta_0s-\delta_{1/2})}\;\;.\label{weak}
\end{equation}

This procedure and similar adiabatic methods led to a general
understanding of the behavior of weakly coupled Josephson junctions.
However, they are not applicable to small circuits
(of diameter $\stackrel{<}{\sim}
1\mu$m) with inductances $\stackrel{<}{\sim}1$
pH. The point is, that for small
inductances the parameter $l^{-1}$ is not longer small,
e.g. $l^{-1}\approx3$ for $L=1$ pH and $I_C=100\;\mu$A.
In this case the term in parenthesis on the r.h.s. of
Eqs. \ref{rsj1} and \ref{rsj2}, resp., dominates
the remaining terms
and it is not possible to derive reduced equations of
the type mentioned above.

In order to deal with small inductances which already
can be realized technologically today,
we developed an alternative systematic
scheme, which is described in the next section.

\section{A perturbation scheme for strong inductive coupling}
At first, we found it convenient to introduce new variables
\begin{equation}\label{deltasigma}
 \Delta=(\phi_2-\phi_1)/2\quad\mbox{and}\quad\Sigma=(\phi_2+\phi_1)/2,
\end{equation}
providing the set of equations
\begin{mathletters}
\begin{eqnarray}
 \dot{\Delta}+\cos\Sigma\sin\Delta&=&\frac{1}{l}(\varphi-2\Delta)
 \label{baseq1},\\
 \dot{\Sigma}+\sin\Sigma\cos\Delta&=&i_0\label{baseq2}.
\end{eqnarray}
\end{mathletters}
These equations already indicate that the behavior of $\Sigma$ is
determined necessarily by the bias current $i_0$ and that of $\Delta$
by the external flux, although coupling makes
things slightly more complicated.

We perform a perturbation expansion valid
for small $l$,
\begin{mathletters}
\begin{eqnarray}\label{pert1}
  \Delta&=&\Delta_0+l\Delta_1+{\cal O}(l^2),\\
  \Sigma&=&\Sigma_0+l\Sigma_1+{\cal O}(l^2).\label{pert2}
\end{eqnarray}
\end{mathletters}
This ansatz resembles the ''slowly-varying amplitude method''
developed several years ago \cite{jain1,thompson1} for
$l^{-1}$ then and resulting in reduced equations similar to
those mentioned above.
Expanding the $\sin$ and $\cos$ terms, it is not sufficient that the
condition
\begin{equation}
 |l\Sigma_1|\ll|\Sigma_0|
\end{equation}
is fulfilled, but additionally we must demand the much more rigid
conditions
\begin{equation}
 |l\Sigma_1|\ll1\quad\mbox{and}\quad|l\Delta_1|\ll1.
\end{equation}
Investigation of the final result shows,
that this can be met for all $\varphi$ by choosing a
sufficiently small $l$. Only the case $\varphi=\pi$ requires a
slightly more involved consideration.

After inserting the expansion (\ref{pert1},b) into Eqs.
\ref{baseq1},b one can compare identical
powers of $l$. To lowest, i.e. $(-1)^{st}$, order one finds
\begin{equation}\label{deltabig0}
\Delta_0=\varphi/2
\end{equation}
without solving any differential equation. This general feature
is valid for higher orders of $l$, too: The variables $\Sigma_n$
must be determined by solving a first-order differential equation
(only to lowest ($0^{th}$) order being non-linear), whereas the
$\Delta$'s can be calculated algebraically. Introducing a once more
rescaled time,
\begin{equation}
 {\tilde s}=s\cdot\cos{(\varphi/2)},
 \quad\quad '=\frac{\mbox{d}}{\mbox{d}{\tilde
 s}},\label{stilde}
\end{equation}
the equation for $\Sigma_0$ (0$^{th}$ order w.r.t. $l$) becomes
very similar to that of an autonomous junction,
\begin{equation}\label{sigma0}
\Sigma'_0+\sin\Sigma_0={\tilde i}_0,\quad
\tilde{i}_0=i_0/\cos(\varphi/2).
\end{equation}
Thus, we can immediately write down the solution,
\begin{equation}\label{sigmabig0}
 \Sigma_0=2\arctan\left(\frac{\overline{\zeta}_0}{i_0+\cos
 (\varphi/2)}
 \tan\Big(\frac{\overline{\zeta}_0s}{2}\Big)\right)+\frac{\pi}{2},
\end{equation}
with
\begin{equation}
 \overline{\zeta}_0=\sqrt{i_0^2-\cos^2(\varphi/2)}\label{zeta}
\end{equation}
where we have imposed the initial condition
\begin{equation}
 \Sigma_0(s=0)=\pi/2.
\end{equation}
Because the $\Delta_n$ are determined algebraically there is no
freedom to specify separate initial conditions for them. To ensure the
validity of the perturbation expansion, it is important to specify all
higher $\Sigma_n$ according to
\begin{equation}
 \Sigma_n(s=0)\;=\;0\quad\mbox{for}\quad n\ge1.\label{initial}
\end{equation}

With $\Delta_0$ and $\Sigma_0$ given $\Delta_1$ can be determined
algebraically,
\begin{equation}
 \Delta_1=\frac{1}{2}\left(\frac{
 \overline{\zeta}_0\sin(\overline{\zeta}_0s)
 }{i_0+\cos(\varphi/2)\cos(\overline{\zeta}_0s)}\sin
 (\varphi/2)\right).\label{delta1}
\end{equation}
This expression automatically satisfies the initial condition
\begin{equation}
 \Delta_1(0)=0
\end{equation}
required for the validity of the
perturbation expansion. For $\Sigma_1$
one obtains the inhomogeneous linear differential equation
\begin{equation}\label{sigma1}
 \dot{\Sigma_1}+\Sigma_1
 \cos\Sigma_0\cos(\varphi/2)-\Delta_1\sin\Sigma_0\sin(\varphi/2)=0,
\end{equation}
admitting the solution
\begin{equation}\label{sigma1sol}
 \Sigma_1=\frac{\tan^2(\varphi/2)}{2(i_0+\cos(\varphi/2)
 \cos\overline{\zeta}_0s)}\left(i_0\cos(\varphi/2)(1-\cos
 \overline{\zeta}_0s)+\overline{\zeta}_0^2\ln\frac{i_0+\cos
 (\varphi/2)\cos\overline{\zeta}_0s}{i_0+\cos(\varphi/2)}
 \right)
\end{equation}
where we already exploited the initial condition, Eq. \ref{initial}.
{}From Eqs. \ref{deltabig0}, \ref{sigmabig0}, \ref{delta1}, and
\ref{sigma1sol} one derives the normalized voltages
\begin{eqnarray}\label{voltages}
&&v_{1/2}=
\frac{\bar{\zeta}_0^2}{i_0+\cos(\varphi/2)\cos\bar{\zeta}_0s}
+\mbox{\large $l$}\frac{\bar{\zeta}_0\sin(\varphi/2)}{2(i_0+\cos(\varphi/2)
\cos\bar{\zeta}_0s)^2}\times\\\hfill\nonumber
\\
&&\Bigg[\sin(\varphi/2)\sin\bar{\zeta}_0s
\Bigg(i_0+\cos(\varphi/2)+
\frac{\bar{\zeta}_0^2}{\cos(\varphi/2)}
\ln\frac{i_0+\cos(\varphi/2)\cos\bar{\zeta}_0s}
{i_0+\cos(\varphi/2)}\Bigg)\nonumber\\
&&\mp\bar{\zeta}_0\Bigg(
\cos(\varphi/2)+i_0\cos\bar{\zeta}_0s\Bigg)\Bigg].\nonumber
\end{eqnarray}

In this way we derived an analytic expression for the voltage
on junction 1 and 2 resp., valid to first order w.r.t. $l$.
It has to be compared with the solution (\ref{weak}) for weakly
coupled elements. Both have in common, that
there always exists a solution showing
phase-locking for all values of the external field. However, contrary
to the case of weak coupling, where the phase shift scales approximately
with external flux, we will shortly see that it is negligibly small
for all values of the external field, except for $\varphi\approx\pi$. On
the other hand, the frequency, being flux-independent for weakly
coupled elements, becomes flux-dependent according to
Eqs. \ref{voltages},10, detuning the cell this way.

So far, we could not derive an equation of motion (a kind of reduced
equation) for the phase shift of strongly coupled
elements directly from the
basic equations as it is possible for weakly coupled elements.
Therefore we must adopt an appropriate alternative
definition of phase difference.
We define phase shift as the difference of the
mean value crossings of the lowest harmonics of $v_1$ and $v_2$
according to (\ref{voltages}).

To proceed, we must evaluate the lowest Fourier
coefficients of Eqs. \ref{voltages},
\begin{equation}\label{fourier}
 v_{1/2}=\frac{\alpha}{2}+a_{1/2}\cos \overline{\zeta}_0s+
 b_{1/2}\sin \overline{\zeta}_0s.
\end{equation}
After some calculation including a Taylor expansion of the
logarithm one obtains
\begin{eqnarray}
 \alpha&=&2\overline{\zeta}_0,\\
 a_{1/2}&=&\frac{\overline{\zeta}_0}{i_0+\overline{\zeta}_0}
 (-2\cos(\varphi/2)\mp
 l\overline{\zeta}_0\sin(\varphi/2))\label{a12},\\
 b_{1/2}&=&b=l\frac{\sin^2(\varphi/2)\cos(\varphi/2)
 }{(i_0+\overline{\zeta}_0)}
 \left(1+\frac{\cos{(\varphi/2)}}{i_0}
 +\frac{\overline{\zeta}_0^3}{4i_0^2(i_0+\overline{\zeta}_0))}\right).
 \label{b}
\end{eqnarray}
Note that the Fourier coefficient
$b$ being proportional to $l$ is small compared to $a_{1/2}$; the
coefficient $a_{1/2}$ itself is dominated by its first term, except
for $\varphi=\pi$. Thus, with the possible exception of the vicinity
of this value,
both voltages are in phase independent of the external flux.

Adopting the definition given before, one can derive a
formula for the phase shift
$\delta$ as a function of the Fourier coefficients,
\begin{equation}\label{deltafourier}
 \delta=\arccos\left(\frac{a_1a_2+b^2}{\sqrt{b^2(a_1^2+a_2^2+b^2)
 +a_1^2a_2^2}}\right),\quad\quad( 0\le\varphi\le\pi).
\end{equation}
Deriving this formula, no assumption has been made about the
structure or order of magnitude of the Fourier coefficients.
Extension to $\pi<\varphi\le2\pi$
needs a more subtle investigation of the solution itself
to get the correct branch of the $\arccos$; in this case
one obtains
\begin{eqnarray}\label{delta2}
  \delta=\mbox{sgn}(\varphi-\pi)\left[
 \pi-\arccos\left(\frac{a_1a_2+b^2}{
 \sqrt{b^2(a_1^2+a_2^2+b^2)
 +a_1^2a_2^2}}\right)\right]+\pi,\\
 (0\le\varphi\le2\pi).\nonumber
\end{eqnarray}
Fig. \ref{fig3} shows the phase shift between $v_1$ and $v_2$
as a function of external flux $\varphi$.

Our analytical approach was accompanied by numerical investigations
exploiting the  Personal Superconducting Circuit ANalyzer program PSCAN
\cite{odintsov1,polonsky1} (Fig. \ref{fig4}).
Comparison of Figs. \ref{fig3} and \ref{fig4} shows that
even for $l=1$ where the analytical approximation is not longer
valid results are quite similar to those of the numerical
simulation. Both figures show that already in this case the behavior is
qualitatively different from that of weakly coupled
elements. Thus, in the intermediate
regime $l\approx1$ the strong coupling scheme provides a better
approximation to the actual behavior of the SQUID cell than the weak
coupling results do.

There are two limiting cases of special interest.
For vanishing external flux the elements behave in the same way as
one junction; the phase shift between voltages vanishes and the
elements oscillate with voltages
\begin{equation}
  v_1=v_2=\frac{\zeta_0^2}{i_0+\cos\zeta_0s}.
\end{equation}
This behavior is plausible, because
in this case there is no current flowing through the shunt and both
elements act essentially as one element. For
$\varphi=\pi$ one obtains
\begin{eqnarray}
  \textstyle v_1&=&i_0
  -\frac{l i_0}{2}\sin i_0s
  +\frac{l}{4}\sin 2i_0s,\\
  \textstyle v_2&=&i_0
  +\frac{l i_0}{2}\sin i_0s
  +\frac{l}{4}\sin 2i_0s.
\end{eqnarray}
This clearly indicates that the phase difference $\delta$
between the voltages according to our definition
is equal to $\pi$.
Within the narrow transition region between the
regimes $\delta\approx0$
and $\delta\approx\pi$ oscillations become highly non-harmonic (cf.
the voltages in Figs. \ref{fig5}a and \ref{fig5}b).
This is caused by higher harmonics becoming dominant in
comparison to the lowest harmonic with frequency $\overline{\zeta}_0$.
In this case the definition of phase difference based on the
lowest harmonic of the Fourier expansion (\ref{fourier})
as well in the analytical calculation as in the numerical procedure
becomes less reliable. The jump observed in the numerical curve
(Fig. \ref{fig4}) may be assigned to this fact.

To derive a more simple rule of thumb
Eq. \ref{deltafourier}
can be rewritten as
\begin{equation}
\delta(\varphi)=\arctan(a_2/b)-\arctan(a_1/b),\quad\quad0\le \varphi<\pi.
\end{equation}
with $a_{1/2}$ and $b$ according to (\ref{a12},\ref{b}).
We consider this formula for very small, but finite $l$. In this case,
the second $\arctan$ approaches the value $-\pi/2$. The first one
changes its sign at the flux $\varphi^*$,
\begin{equation}
 \cos(\varphi^*/2)-\frac{l}{2}\overline{\zeta}_0\sin(\varphi^*/2)=0.
\end{equation}
Exploiting this formula while neglecting
higher orders in $l$, one obtains
\begin{equation}
 \varphi^*\approx \pi-i_0l.
\end{equation}
This provides a simple approximation for the phase shift of the cells
under investigation,
\begin{equation}
\delta\approx\pi\theta(\varphi-\varphi^*)\quad (\varphi\le\pi).
\end{equation}
Fig. \ref{fig6} shows that for sufficiently small $l$
the solution is indeed
perfectly approximated by a Heaviside step function. This approximation
might be useful considering more complicated arrays.

Another quantity of interest is the $I$-$V$ characteristics of the cells
under investigation. From (\ref{fourier}) one easily obtains
\begin{equation}
 \overline{v}_{1/2}=\sqrt{i_0^2-\cos^2(\varphi/2)}.
\end{equation}
This reproduces a well-known result\cite{likharev1}:
The $I$-$V$ characteristics of a
small-inductance SQUID has a hyperbolic shape, the vertex being
dependent on the external flux.

\section{Parameter-splitting in strongly coupled cells}
\label{splitting}
Real junctions never have identical parameters.
The response to parameter differences becomes particularly
important in large arrays, where one usually cannot avoid a
parameter-splitting of the order of one p.c., at least. In this
section we consider junctions having different critical currents
as well as normal resistances,
\begin{equation}
 I_{C_1}\ne I_{C_2},\quad R_{N_1}\ne R_{N_2},
\end{equation}
with the subsidiary condition
\begin{equation}
 I_{C_1}R_{N_1}=I_{C_2}R_{N_2},
\end{equation}
which is usually realized as a consequence of the technological process
with a good accuracy. Introducing the mean
critical current
\begin{equation}
 I_C=\frac{1}{2}(I_{C_1}+I_{C_2})
\end{equation}
and the parameter-splitting
\begin{equation}
 \vartheta=\frac{I_{C_2}-I_{C_1}}{I_{C_2}+I_{C_1}},
\end{equation}
one derives the following RSJ-model equations for the cells shown in
Fig. \ref{fig1}:
\begin{mathletters}
\begin{eqnarray}
 &&\dot{\phi_1}+\sin\phi_1=\frac{i_0}{1-\vartheta}-
 \frac{1}{l(1-\vartheta)}(\phi_1-\phi_2+\varphi),\label{rsj1param}\\
 &&\dot{\phi_2}+\sin\phi_2=\frac{i_0}{1+\vartheta}+
 \frac{1}{l(1+\vartheta)}(\phi_1-\phi_2+\varphi)\label{rsj2param}.
\end{eqnarray}
\end{mathletters}

As before, it is advantageous to introduce new variables $\Delta$
and $\Sigma$ according to Eq. \ref{deltasigma}. Eqs.
\ref{baseq1},b are then modified to
\begin{mathletters}
\begin{eqnarray}
 \dot{\Delta}+\cos\Sigma\sin\Delta&=&-
 \frac{\vartheta}{1-\vartheta^2}\,i_0+
 \frac{1}{l(1-\vartheta^2)}(\varphi-2\Delta),
 \label{baseq1param}\\
 \dot{\Sigma}+\sin\Sigma\cos\Delta&=&\frac{1}{1-\vartheta^2}\,i_0-
 \frac{\vartheta}{l(1-\vartheta^2)}(\varphi-2\Delta).
 \label{baseq2param}
\end{eqnarray}
\end{mathletters}
Some effects are already qualitatively displayed by this
couple of equations.
(i) To first order in $\vartheta$ there is a
correction of the magnetic flux $\sim-i_0l\vartheta$. (ii) There is a
correction of the bias current
$\vartheta(\varphi-2\Delta)/l$ being of
first order, too. It includes an additional coupling via $\Delta$.
Eqs. (\ref{baseq1param},b) indicate that for
weak coupling $(l\gg1, \vartheta\ll1)$ the additional magnetic flux
dominates, an effect which has already been observed (cf. Eq.
13.30b in\cite{likharev1}). However, for strong coupling
$(l\ll1, \vartheta\ll1)$ this term is of second order only.
It turns out that the difference $\varphi-2\Delta$ is of the
order of $l$, so the addition to the bias current is of first order in
$\vartheta$ and dominates.

First of all, we are interested in the maximum parameter-splitting
which is possible without destroying synchronization.
For this purpose, the splitting-parameter
$\vartheta$ should not be considered small from the beginning.
As before, we perform a perturbation expansion w.r.t. $l$ according to
(\ref{pert1},\ref{pert2}). To lowest, i.e. $-1^{st}$, order we again
obtain $\Delta_0=\varphi/2$. To $0^{th}$ order, we get the system of
equations\begin{mathletters}
\begin{eqnarray}\label{basedel1}
 \dot{\Sigma}_0+\sin\Sigma_0\cos(\varphi/2)&=&
 \frac{i_0}{1-\vartheta^2}
 +\frac{2\vartheta}{1-\vartheta^2}\Delta_1,\\
 \cos\Sigma_0\sin\Delta_0&=&-\frac{\vartheta}{1-\vartheta^2}i_0-
 \frac{2}{1-\vartheta^2}\Delta_1.\label{basedel2}
\end{eqnarray}
\end{mathletters}
Again, $\Sigma_0$ has to be determined by solving a differential
equation, whereas $\Delta_1$ is calculated algebraically.
The new feature is an additional coupling between both variables
caused by the last term on the r.h.s. of Eq. \ref{basedel1}.
Combining both equations, one obtains
\begin{equation}\label{sigma0split}
 \dot{\Sigma}_0+\sin\Sigma_0\cos(\varphi/2)+\vartheta\cos\Sigma_0
 \sin(\varphi/2)=\frac{i_0}{1-\vartheta^2}.
\end{equation}
In comparison to (\ref{sigma0}) this equation shows an additional
non-linearity due to the parameter-splitting. Moreover, the current
is divided by $(1-\vartheta^2)$, indicating
that a splitting of critical current can lower the
current necessary for the onset of oscillations.

Eq. \ref{sigma0split} can be handled exactly.
There are four different types of solutions \cite{gradstein1}.
Only one of them shows the continuous transition to
the case $\vartheta=0$ and the corresponding
oscillating voltage: It is realized for
\begin{equation}
 \frac{i_0^2}{1-\vartheta^2}>\vartheta^2\sin^2(\varphi/2)+
 \cos^2(\varphi/2).
\end{equation}
Further estimation gives the bound for oscillations to occur,
\begin{equation}
 i_0^2>1-\vartheta^2.
\end{equation}
For $\vartheta=0$ this reproduces a well-known fact. In addition, it
proves the conjecture above that growing parameter-splitting leads to
an effective lowering of the critical current.

With this condition fulfilled we could evaluate $\Sigma_0$ and from
(\ref{basedel2}) $\Delta_1$. However, although
$\Sigma_1$ has to be determined from a first-order
linear differential equation, the resulting integrals are rather
intricate.
Thus, we performed a perturbative treatment of the system
(\ref{baseq1param},b) not only w.r.t. $l$, but w.r.t.
$\vartheta$, too. This is more delicate, of course, because there
are two parameters involved.
To discuss the $l$- and $\vartheta$-dependence
independently, its not wise to specify
the ratio $l/\vartheta$ from the beginning. We only suppose
$l\ll1$ and $\vartheta\ll1$, leaving the ratio $\vartheta/l$
unspecified. To first order, we write down the expansion
\begin{mathletters}
\begin{eqnarray}
 \Delta&=&\Delta_0+l\Delta_{10}+\vartheta\Delta_{01},\\
 \Sigma&=&\Sigma_0+l\Sigma_{10}+\vartheta\Sigma_{01}.
\end{eqnarray}
\end{mathletters}
Inserting into (\ref{baseq1param},b)
and comparing equal
orders $l^m\vartheta^n$,  one obtains the set of equations
necessary to evaluate the $\Delta$'s and $\Sigma$'s. For
$\Delta_0,\Sigma_0,\Delta_{10}$, and $\Sigma_{10}$
one obtains similar
equations as before.
Furthermore, one observes $\Delta_{01}=0$; thus, no
additional phase shift is caused by the parameter-splitting. For
$\Sigma_{01}$ one obtains an equation similar in structure to
that for $\Sigma_{10}$ (cf. Eq. \ref{sigma1}),
\begin{equation}
 \dot{\Sigma}_{01}+\Sigma_{01}\cos\Sigma_0\cos(\varphi/2)
 +\cos\Sigma_0\sin(\varphi/2)=0.
\end{equation}
It admits the solution
\begin{equation}
 \Sigma_{01}=\frac{1-\cos\overline{
 \zeta}_0s}{i_0+\cos(\varphi/2)\cos\overline{
 \zeta}_0s}\sin(\varphi/2).
\end{equation}
Thus, weak parameter-splitting leads to an
additional in-phase contribution
\begin{equation}
 \vartheta(i_0+\cos(\varphi/2))\sin\overline{\zeta}_0s
 \frac{\overline{\zeta}_0s\sin(\varphi/2)}{(i_0-\cos
 (\varphi/2)\cos\overline{\zeta}_0s)^2}
\end{equation}
to be added to the voltages (\ref{voltages}). The Fourier coefficients
$a_{1/2}$  are unaffected by this, whereas there is an additional
contribution to $b_{1/2}$,
\begin{equation}
 b_{1/2}^{\vartheta}=2\vartheta\Big(i_0+\cos(\varphi/2)\Big)
 \frac{\sin(\varphi/2)}{i_0+\overline{\zeta}_0}.
\end{equation}

The solution obtained this way proves our earlier conjecture on the
dominant contribution in the strong-coupling case (Fig. \ref{fig7}a).
One observes that the phase shift, being slightly raised
generally is considerably lowered for $\varphi=\pi$.
To lowest order of our analytic approximation (valid for
strong coupling and weak parameter splitting) there is
no indication of a shift of the peak caused by the parameter splitting.
This is confirmed by comparison with numerical simulation,
as long as parameter splitting
is sufficiently small ($\vartheta \stackrel{<}{\sim}0.2$).
For larger $\vartheta$ the numerical result (Fig. \ref{fig7}b)
gives a first hint to the peak shift.

\section{Capacitively shunted junctions}
The influence of the displacement current flowing through
the junctions was neglected up to now. This is justified,
as long as the McCumber
parameter\cite{mccumber1,steward1}
\begin{equation}
 \beta=\frac{2e}{\hbar}I_CR_N^2C
\end{equation}
is negligible.

In this section, we will investigate the influence of $\beta\ne0$.
The displacement current adds a second-derivative term to the RSJ
model equations (sometimes called RCSJ model equations then),
\begin{mathletters}
\begin{eqnarray}
 \beta\ddot{\phi_1}+\dot{\phi_1}+\sin\phi_1&=&i_0
 -\l^{-1}(\phi_1-\phi_2+\varphi),\label{rcsj1}\\
 \beta\ddot{\phi_2}+\dot{\phi_2}+\sin\phi_2&=&i_0
 +\l^{-1}(\phi_1-\phi_2+\varphi)\label{rcsj2}.
\end{eqnarray}
\end{mathletters}
In general, the second derivative may change the character of the
differential equations completely; for instance, it is well-known
that there appear new types of solutions showing chaotic
behavior\cite{marcus1,miracky1}. Here, we will restrict our
treatment to small $\beta$ ($\beta\ll1$)
guaranteeing a continuous transition to
the former solution for $\beta=0$.

Again, it is recommended to combine Eqs.
\ref{rcsj1},b obtaining
\begin{mathletters}
\begin{eqnarray}
 \beta\ddot{\Delta}+\dot{\Delta}+
 \cos\Sigma\sin\Delta&=&\frac{1}{l}(\varphi-2\Delta)
 \label{baseqbeta1},\\
 \beta\ddot{\Sigma}+\dot{\Sigma}+
 \sin\Sigma\cos\Delta&=&i_0\label{baseqbeta2}.
\end{eqnarray}
\end{mathletters}
One clearly sees, that both equations are affected by the additional
$\beta$-terms. Again, both $l$ and $\beta$ are supposed to be small
parameters justifying the expansion
\begin{mathletters}
\begin{eqnarray}
 \Delta&=&\Delta_0+l\Delta_{10}+\beta\Delta_{01},\\
 \Sigma&=&\Sigma_0+l\Sigma_{10}+\beta\Sigma_{01}.
\end{eqnarray}
\end{mathletters}
The resulting equations for $\Delta_0,\Sigma_0,\Delta_{10},$ and
$\Sigma_{10}$, resp., are essentially the same as before. For
$\Delta_{01}$ one readily recovers
\begin{equation}
 \Delta_{01}=0.
\end{equation}
The only new equation concerns $\Sigma_{01}$,
\begin{mathletters}
\begin{eqnarray}
 \dot{\Sigma}_{01}+
 \Sigma_{01}\cos\Sigma_0\cos{(\varphi/2)}=-\ddot{\Sigma}_{0},
\end{eqnarray}
\end{mathletters}
where we have already exploited some of the previous results. Again,
this is an inhomogeneous linear differential equation, with the
inhomogeneity being determined by the already well-known $\Sigma_0$.
The solution, obeying the correct boundary condition
($\Sigma_{01}(s=0)=0$), is
\begin{equation}\label{sigma01beta}
 \Sigma_{01}=\frac{\overline{\zeta}_0^2}{i_0-\cos(\varphi/2)
 \cos\overline{\zeta}_0s}\ln\frac{i_0+\cos(\varphi/2)
 \cos\overline{\zeta}_0s}{i_0+\cos(\varphi/2)}.
\end{equation}
One obtains a logarithmic structure similar to that observed
earlier in formula
(\ref{sigma1sol}). The term (\ref{sigma01beta}) provides
to both voltages a contribution
\begin{equation}
 \dot{\Sigma}_{01}=\frac{\overline{\zeta}_0^3\cos
 (\varphi/2)\sin\overline{\zeta}_0s}{(i_0+\cos(\varphi/2)
 \cos\overline{\zeta}_0s)^2}\left(\ln\frac{i_0+\cos
 (\varphi/2)\cos\overline{\zeta}_0s}{i_0+\cos(\varphi/2)}-1
 \right).
\end{equation}
Because the logarithmic structure is already present in
(\ref{voltages}), it is not hard to evaluate the corresponding
capacitive contribution to be added to
the Fourier coefficients $b_{1/2}$ according to (\ref{b}),
\begin{equation}\label{bbeta}
 b_{1/2}^{\beta}=b^{\beta}=
 -2\beta\overline{\zeta}_0^2\frac{\cos(\varphi/2)}{(i_0+\overline{
 \zeta}_0)}\left(1+\frac{\cos(\varphi/2)}{i_0}
 -\frac{\overline{\zeta}_0\cos^2(\varphi/2)}{4
 i_0^2(i_0+\overline{\zeta}_0)}\right).
\end{equation}

The phase difference obtained by inserting (\ref{b}) and
(\ref{bbeta}) together with the unchanged coefficients (\ref{a12})
into (\ref{delta2}) is shown in Fig. \ref{fig8}a.
Results of a numerical calculation performed in parallel
are given in Fig. \ref{fig8}b. The
general tendency is that a
non-vanishing capacitance $(\beta\stackrel{<}{\sim}1)$
slightly enhances the phase shift without qualitatively changing the
general behavior. For $\beta>0.5$ the agreement between analytical
approximation and numerical simulation becomes worse, but the
same general tendency is still preserved.
This is, of course, simply a result of the fact
that higher orders in $\beta$ are no longer negligible.

\section{Strongly coupled SQUID cells under microwave radiation}
\label{micro}
There is some interest in the behavior of the SQUID cells
under microwave radiation from at least two points of view.
First of all, the topic
is interesting for the construction of sensitive microwave detectors.
Secondly, knowledge of the behavior under microwave radiation
is necessary for the study of synchronization in larger arrays, where
the long-range interaction via external shunts acts
similar to an external microwave signal.

The external microwave signal can be described by an additional ac
current, leading to the system of equations
\begin{mathletters}
\begin{eqnarray}
 \dot{\phi_1}+\sin\phi_1&=&i_0-\l^{-1}(\phi_1-
\phi_2+\varphi)+i_{\omega}\sin\omega s,\\
 \dot{\phi_2}+\sin\phi_2&=&i_0+\l^{-1}(\phi_1-
\phi_2+\varphi)+i_{\omega}\sin\omega s.
\end{eqnarray}
\end{mathletters}
As a result,
only the equation for the sum variable $\Sigma$ is affected and becomes
\begin{equation}
 \dot{\Sigma}+\sin\Sigma\cos\Delta=i_0+i_{\omega}\sin\omega s,
\end{equation}
whereas Eq. \ref{baseq1} for the difference variable $\Delta$
remains unchanged.
We apply a perturbation scheme similar to that used before.
{}From the beginning we will assume $l\ll1$, as before, justifying the
expansion (\ref{pert1},b).
Solving for $\Sigma_0$ and $\Sigma_1$, resp., we must
take $i_{\omega}$ to be small in some intermediate steps, too.

The first result is
\begin{equation}
 \Delta_0=\varphi/2,
\end{equation}
as usual. The corresponding equation for $\Sigma_0$,
\begin{equation}\label{sigma0ext}
 \dot{\Sigma}_0+\sin\Sigma_0\cos(\varphi/2)=
 i_0+i_{\omega}\sin\omega s,
\end{equation}
is decisive for the behavior of the solution. Introducing the scaled
time $\tilde{s}$ according to (\ref{stilde}), we obtain
\begin{equation}
 \Sigma'_0+\sin\Sigma_0=\tilde{i}_0+\tilde{i}_{\omega}
 \sin\tilde{\omega}\tilde{s},\quad\quad \tilde{i}_{\omega}=
 \frac{i_{\omega}}{\cos(\varphi/2)},\quad\quad\tilde{\omega}=
 \frac{\omega}{\cos(\varphi/2)},
\end{equation}
which formally has the
same structure as the equation describing an autonomous Josephson
junction under external irradiation\cite{krech8,thompson1,likharev3}.

It is well-known that phase-locking
of an autonomous junction takes place only if
the frequency of the external microwave does not deviate too
far from the inherent Josephson frequency $\zeta_0=\sqrt{i_0^2-1}$,
\begin{equation}
 \left|\frac{2\zeta_0}{i_{\omega}}(\omega-\zeta_0)\right|<1\quad\quad
 \mbox{(autonomous contact)}.
\end{equation}
The main new feature in our case is, that the corresponding quantities
substituted for $i_0,i_{\omega},$ and $\omega$ resp.
according to
\begin{eqnarray}
 i_0&\rightarrow&\tilde{i}_0,\\
 i_{\omega}&\rightarrow&\tilde{i}_{\omega},\\
 \omega&\rightarrow&\tilde{\omega}
\end{eqnarray}
are dependent on the external magnetic flux. Exploiting the
corresponding equation
\begin{equation}
 \left|\frac{2\tilde{\zeta}_0}{\tilde{i}_{
 \omega}}(\tilde{\omega}-\tilde{\zeta}_0)\right|<1,
\end{equation}
one obtains the phase-locking condition
\begin{equation}
 \overline{\zeta}_0-\frac{i_{\omega}
 \cos(\varphi/2)}{2\overline{\zeta}_0}\le
 \omega\le\overline{\zeta}_0+\frac{i_{\omega}
 \cos(\varphi/2)}{2\overline{\zeta}_0}.\label{synchcond}
\end{equation}
An interesting conclusion from Eq. \ref{synchcond} is that for
external flux $\varphi=\pi$
the range of phase-locking shrinks to a single point,
$\omega=\overline{\zeta}_0=i_0$.
In this case, for all practical purposes
phase-locking disappears at all.
This is confirmed by Fig. \ref{fig9}, showing the range
of phase-locking against external flux $\varphi$.
The reason for this behavior is obvious from examining
(\ref{sigma0ext}): For $\varphi=\pi$ the non-linear term vanishes thus
removing the non-linearity of the equation at all.
One more observation is that the center of the locking-range
determined by $\overline{\zeta}_0$ becomes flux-dependent, too.

In case of
phase-locking, phase shift between the external microwave
and the circuit (the last being
characterized by the sum variable $\Sigma$)
can be deduced from comparison with the autonomous contact yielding
\begin{equation}
 \delta_0=\arcsin\frac{2\overline{\zeta}_0(\omega-\overline{
 \zeta}_0)}{i_{\omega}\cos(\varphi/2)}.\label{phaseext}
\end{equation}

To verify our result in an independent way, we performed a numerical
operation range analysis automatically
integrating the corresponding
differential equations and checking, whether the results lie within a
certain bound. The output
of this analysis indicated by diamonds in Fig. \ref{fig9}
is in excellent agreement with the analytical results.
In view of the experimental setup
the figure should be interpreted as follows: For a fixed bias current
and frequency of the external microwave radiation there is a
limited range of flux indicated in Fig. \ref{fig9},
within which synchronization occurs.
Within this range the whole cell oscillates with the
microwave frequency, $\omega$. For small microwave intensities this
puts rather severe conditions on the external flux,
as is observed by comparing Figs. \ref{fig9}a, b and c.

In case of phase-locking, we obtain to $0^{th}$ order
of perturbation theory
\begin{equation}
 \Sigma_0=2\arctan\left(\frac{\overline{\zeta}_0}{i_0+\cos
 (\varphi/2)}
 \tan\Big(\frac{\omega s}{2}-\frac{\delta_0}{2}\Big)
 \right)+\frac{\pi}{2}
\end{equation}
with the lowest-order phase shift $\delta_0$ according to
(\ref{phaseext}).

Within the next perturbative order, $\Delta_1$ is determined
algebraically as before. The solution is identical to
Eq. \ref{delta1}, the only difference being, that within the
time-dependent arguments one has to substitute
\begin{equation}\label{subst}
 \overline{\zeta}_0s\rightarrow\omega s-\delta_0.
\end{equation}
For $\Sigma_1$ we obtain the equation
\begin{equation}\label{sigma1ext}
 \dot{\Sigma_1}+\Sigma_1\cos\Sigma_0\cos(\varphi/2)-
 \Delta_1\sin\Sigma_0\sin(\varphi/2)=0.
\end{equation}
Substituting $\Sigma_0$ we exploit (\ref{sigma0ext})
neglecting the external current,
bearing in mind that $i_{\omega}$ is small and $\Sigma_1$ is
already of first order. Within this approximation, the solution
has the same general structure as
(\ref{sigma1sol}), where we again have to substitute
(\ref{subst}). As a result, in addition to the
flux-dependent phase shift between the
SQUID circuit voltage oscillations and the external
microwave signal we obtain the same (mostly negligible) phase splitting
between the junction voltages than without radiation.

To summarize, solution (\ref{voltages}) is reproduced with
the only substitution (\ref{subst}).
This has several consequences: (i) The frequency of the
oscillations is determined by the microwave frequency only and turns out
to be independent of external flux within the locking-range.
(ii) If external flux is present in addition to the
microwave radiation, this flux will
limit the range of phase-locking in general
and destroy synchronization at all in case of $\varphi=\pi$.
(iii)  The relative phase of both junctions
is not influenced by the external
radiation up to first order in perturbation theory w.r.t. $l$.
(iv)  There is an additional shift of both
phases relative to the external radiation
according to Eq. \ref{phaseext}, which is controlled by external
flux as well.

\section{Summary}
We investigated the synchronization behavior of three
similar 2-junction SQUID cells with strong inductive coupling.
For this purpose we developed a perturbation scheme appropriate for
small but non-vanishing inductances. The perturbation ansatz
itself is in a certain sense similar
to the ''slowly-varying amplitude method''
developed several years ago. However,
application to strong coupling completely changes the character of the
expansion. Generally, the procedure is more involved than in the case
of weak coupling. Therefore
we were not able to derive an explicit equation of motion for the
phase difference between voltages. In view of this fact, we determined
voltage phase shift from the lowest Fourier coefficients of the
voltages.

For identical junctions without
capacitive shunting we find that for every
value of the external flux a phase-locking between oscillating
voltages takes place like in weakly coupled elements.
However, contrary to the case of weak coupling, the
phase shift is negligibly small
for almost all values of external flux, with the exception of
the vicinity of $\varphi=\pi$. On the other hand, the frequency, being
flux-independent for weakly coupled elements, becomes strongly
flux-dependent in strongly coupled elements and, consequently,
the corresponding $I$-$V$ characteristics too.
The results obtained are compared with numerical calculations.
Generally, a good agreement is observed. Especially it is found, that
the strong-coupling approach
provides a good approximation not only for very small inductances $l$,
but is better suited to describe the intermediate range $l\approx1$
than the ordinary weak-coupling approach.

If both junctions are not identical the influence
of parameter splitting is found to be qualitatively different for
weak and for strong coupling. For weak coupling, parameter splitting
mainly leads to an additional contribution to
the external flux; as a result, the whole
phase-flux-dependence is shifted by some value. In case of
strong coupling, this effect can be neglected and the leading
contribution is a correction of the bias current. This correction acts
in favor of synchronization and lowers the phase shift present in a
small range around $\varphi=\pi$.

In case of identical junctions having a small, but non-vanishing
capacitance the main result is a slight enhancement of the phase
shift, although the qualitative picture is not changed, at least if
$\beta<1$. For $\beta>0.5$ the agreement of analytical results
and numerical simulation becomes less convincing, obviously
showing the
limitations of applicability of the analytical perturbative method. We
should mention that in the case $\beta\ne0$ as well as for
parameter splitting two independent expansion parameters must be
considered small.

Finally, we investigated the behavior of the cells under external
microwave radiation. In this case, we observed a limited
locking range similar to that of an autonomous Josephson
junction under external radiation. However, for a strongly coupled
cell the synchronization range is flux-dependent
and shrinks to zero for $\varphi=\pi$. In addition, the width of
the synchronization range depends on the amplitude of the external
radiation.

Contrary to our results for a freely oscillating cell,
external radiation synchronizes the cell in such a way that the
oscillation frequency becomes flux-independent within the flux-dependent
locking range.
However, within the synchronization range, an additional
phase shift between external radiation and internal oscillations
takes place as well as a shift of the whole synchronization range.

{}From our study one can draw the general conclusion, that two
strongly inductive coupled junctions behave like a free
junction if no flux is present within the cell. External
flux tends to shift the phases between the voltage
oscillations of the two junctions,
but for most practical applications this splitting is negligibly small.
More serious is the fact that already in a cell of identical junctions
the oscillation frequency itself becomes field-dependent.
This detuning of the cell
has several consequences for the construction of
larger arrays. First of all,
external fluxes must be shielded, preferably
by an external superconducting ground plane.
Secondly, additional
fluxes are produced by the array itself, which might seriously
disturb the synchronization \cite{phillips1}.

A possible way to circumvent this problem and to
obtain phase-locking could be to include an external long-range
interaction via an additional
shunt. The methods developed in Sec. \ref{micro} could be helpful in
such an investigation.
For instance, according to our observations,
an external shunt with sufficiently strong coupling-strength
to make the synchronization frequency flux-independent may play a crucial
role for obtaining phase-locking
in large two-dimensional arrays.

\section*{acknowledgments}
This work was supported by a project of the BMBW under contract \# 13N6132
and by DFG under contract \# KR1172/4-1.
The authors would like to express their thanks to BMBW, DFG and DAAD for
financial support. In addition, the authors would like to
thank A. Nowack for a critical reading of the manuscript.
\eject

\begin{figure}
\caption{Three SQUID Cells which can be described with the
strong coupling method described in this paper.}
\label{fig1}
\end{figure}

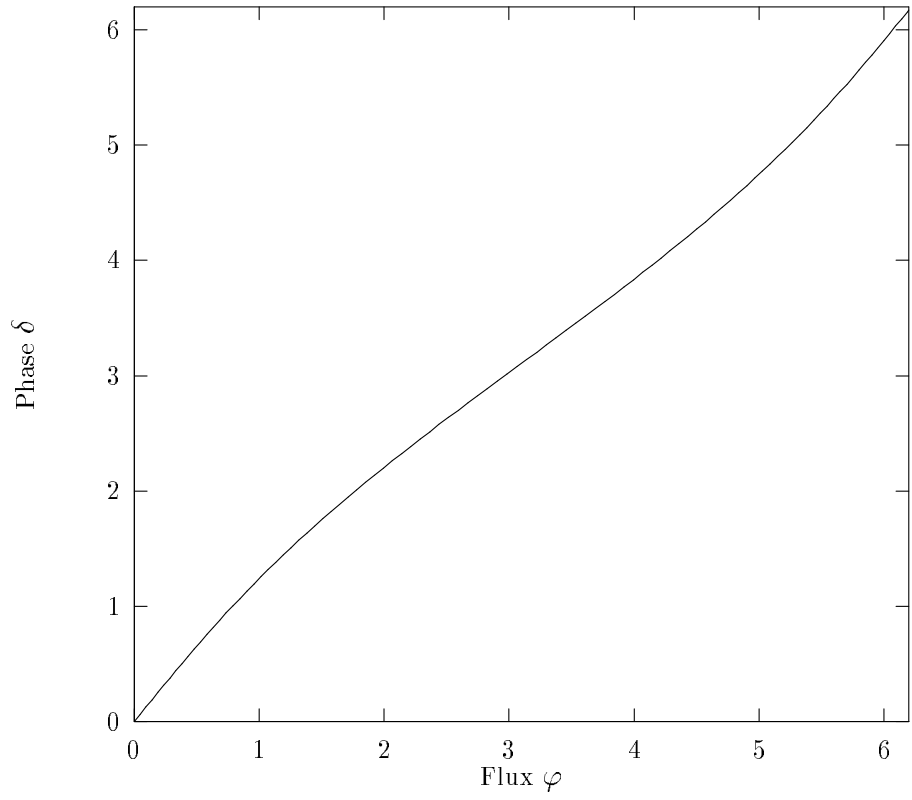
\begin{figure}
\caption{Mean voltage phase shift $\delta$ against
normalized external flux $\varphi$ from analytical approximation
for weak inductive coupling ($i_0=1.5$).}
\label{fig2}
\end{figure}

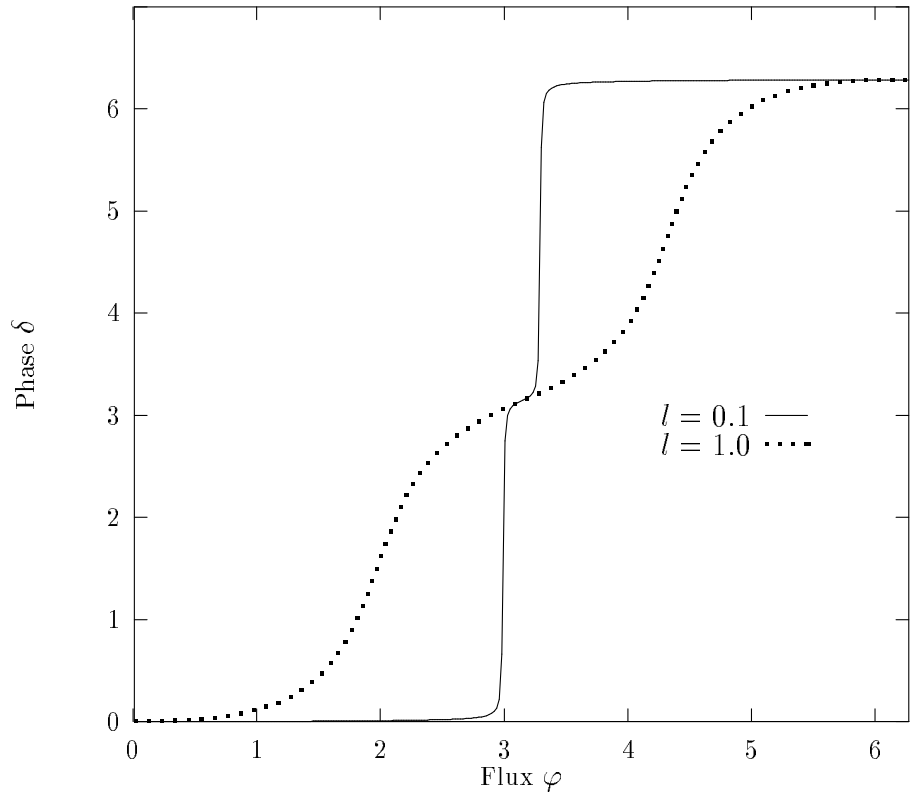
\begin{figure}
\caption{Phase shift $\delta$ against
normalized external flux $\varphi$ for strong inductive coupling
$l=0.1$ and medium inductive coupling $l=1.0$
obtained from analytical approximation
({\protect\ref{delta2}})
($i_0=1.5$).
}
\label{fig3}
\end{figure}

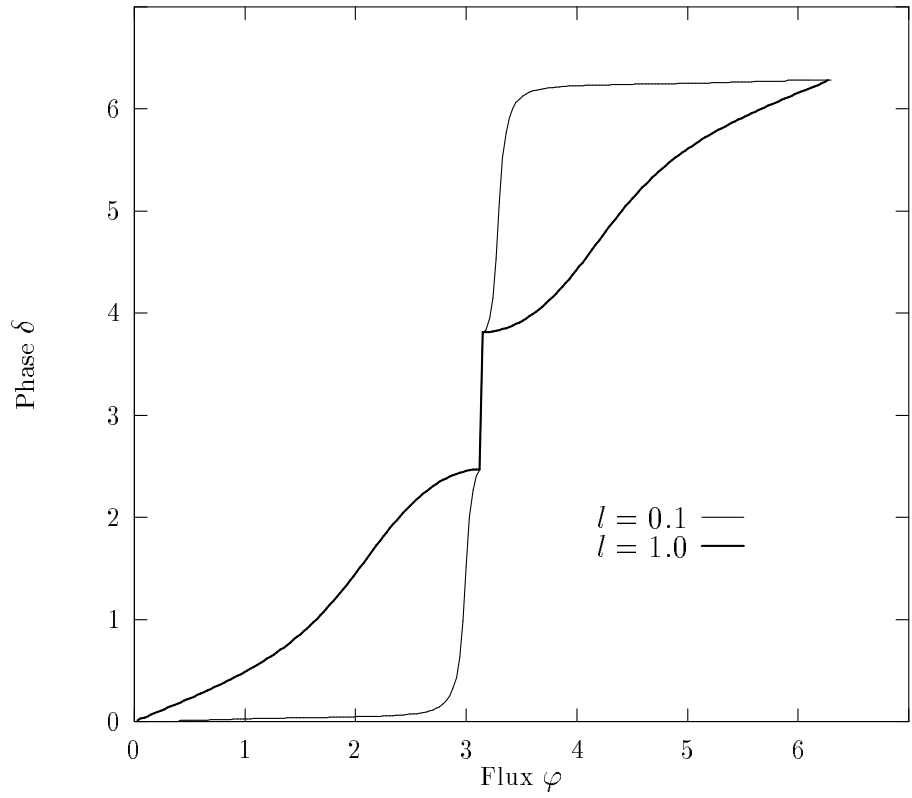
\begin{figure}
\caption{Like Fig. \protect\ref{fig3}, results obtained from
numerical simulation. (A tiny shunt
capacitance $\beta=0.01$ has to be added here.)}
\label{fig4}
\end{figure}

\begin{figure}
\caption{Time-dependence of the voltage
for two different values of the external flux,
a) $\varphi=\pi/2$ b) $\varphi=\pi$, Parameters: $i_0=1.5,\;l=0.1$.}
\label{fig5}
\end{figure}

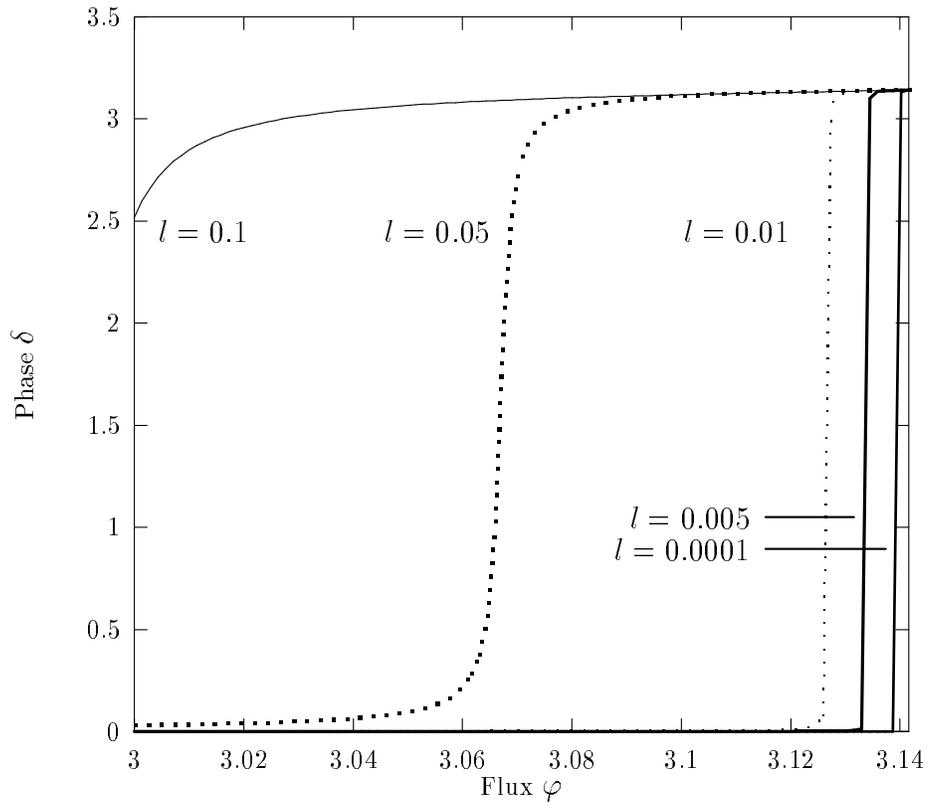
\begin{figure}
\caption{Phase shift $\delta$ against
normalized external flux $\varphi$ for extremely
strong coupling ($0.001\le l\le 0.1$),
obtained from analytical approximation (\protect\ref{delta2})
($i_0=1.5$).
}
\label{fig6}
\end{figure}

\begin{figure}
\caption{The influence of parameter splitting on the phase shift $\delta$
against normalized external flux $\varphi$ for strong coupling
obtained from a) analytical approximation (\protect\ref{delta2}), b)
numerical simulation ($i_0=1.5,\;l=0.1,\;2\le\varphi\le4$).}
\label{fig7}
\end{figure}

\begin{figure}
\caption{The influence of a non-vanishing
capacitive shunt of the Josephson
junctions on the phase difference for $i_0=1.5,\;l=0.1,\;
\beta=0{.}2,\,0{.5},\,1{.}0$. a) analytical approximation, b)
numerical simulation ($i_0=1.5,\;l=0.1,\;2\le\varphi\le4$).
}
\label{fig8}
\end{figure}

\begin{figure}
\caption{Synchronization range for a strongly coupled SQUID
cell under external microwave radiation with frequency
$\omega$ for a)$i_{\omega}=0.1$, b)$i_{\omega}=0.2$,
c)$i_{\omega}=0.4$ (bias current $i_0=1.5$ ).
The diamonds indicate results from numerical
operation range analysis.
}
\label{fig9}
\end{figure}

\bibliographystyle{prsty}


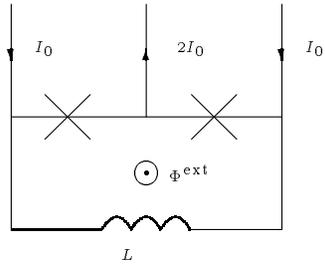
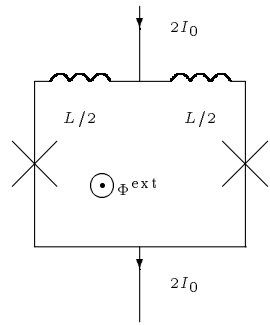
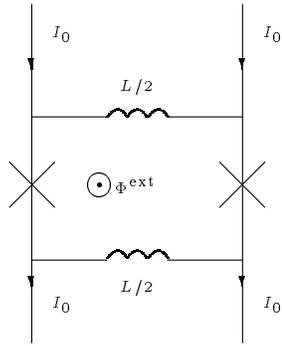
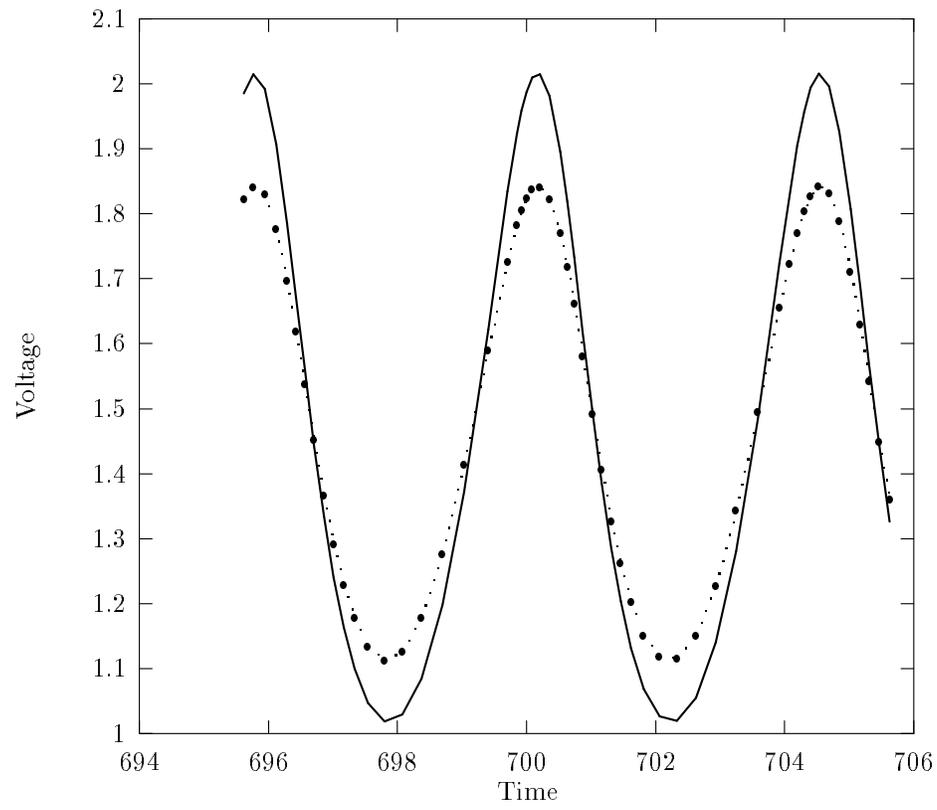
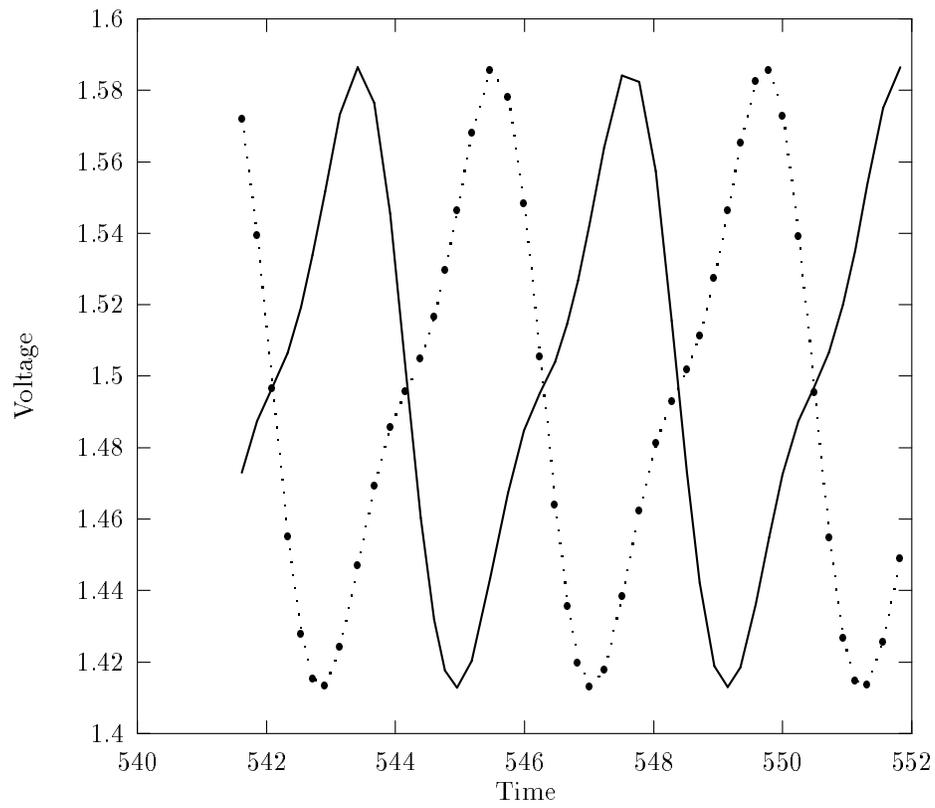
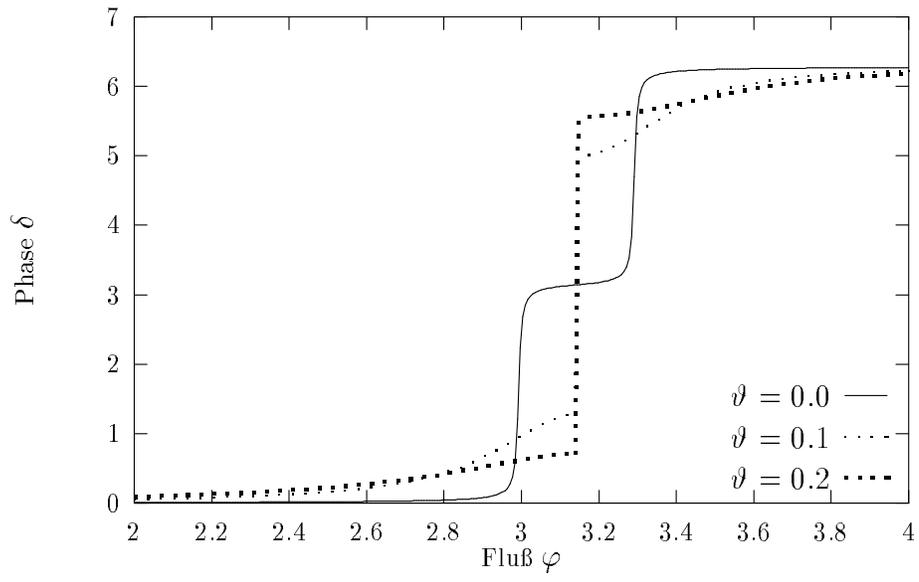
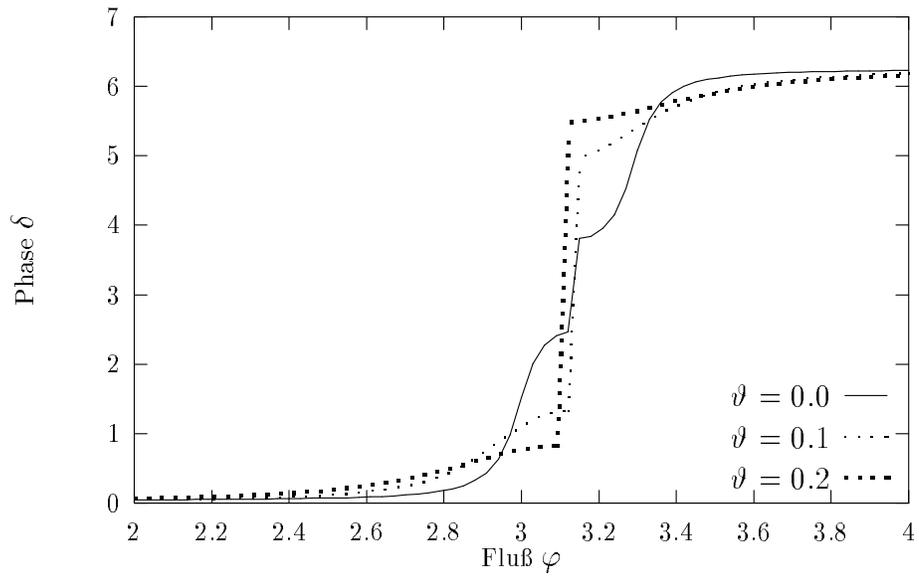
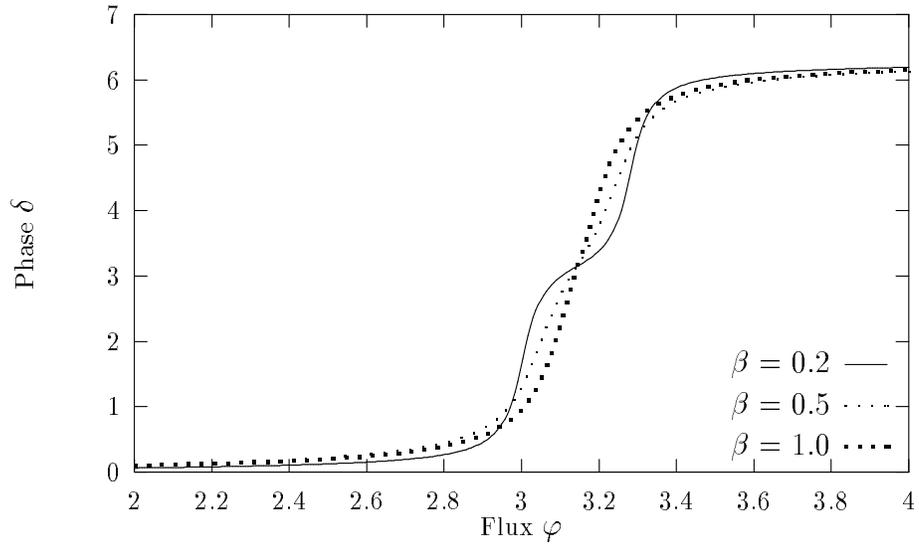
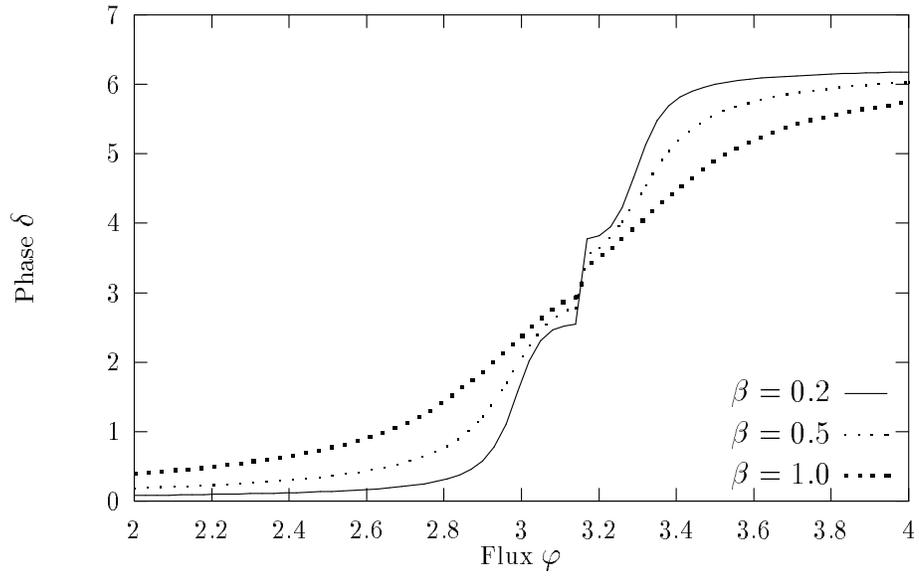
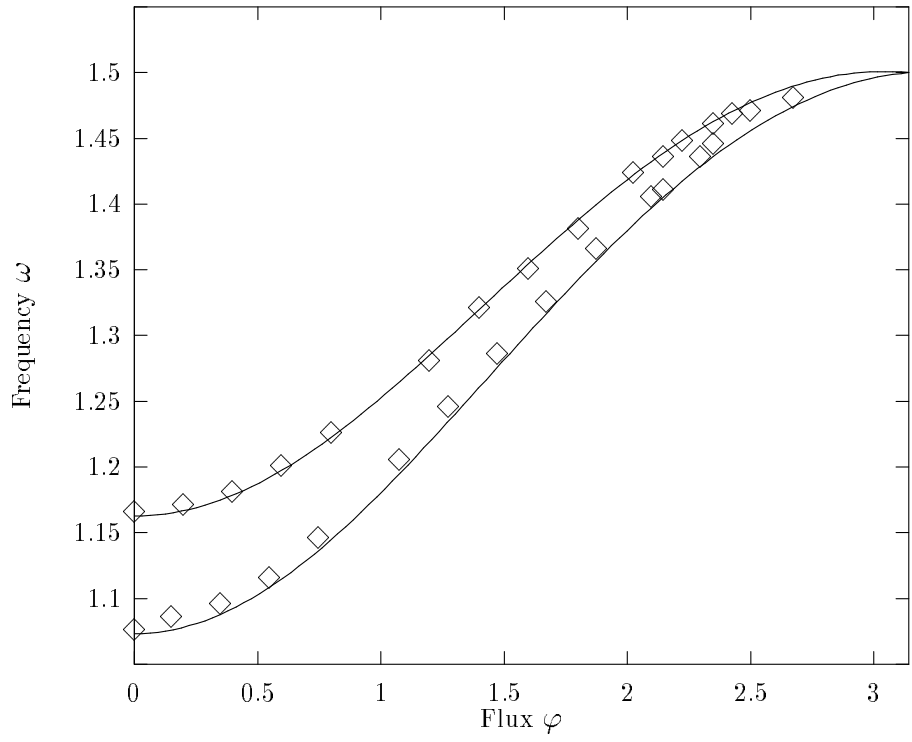
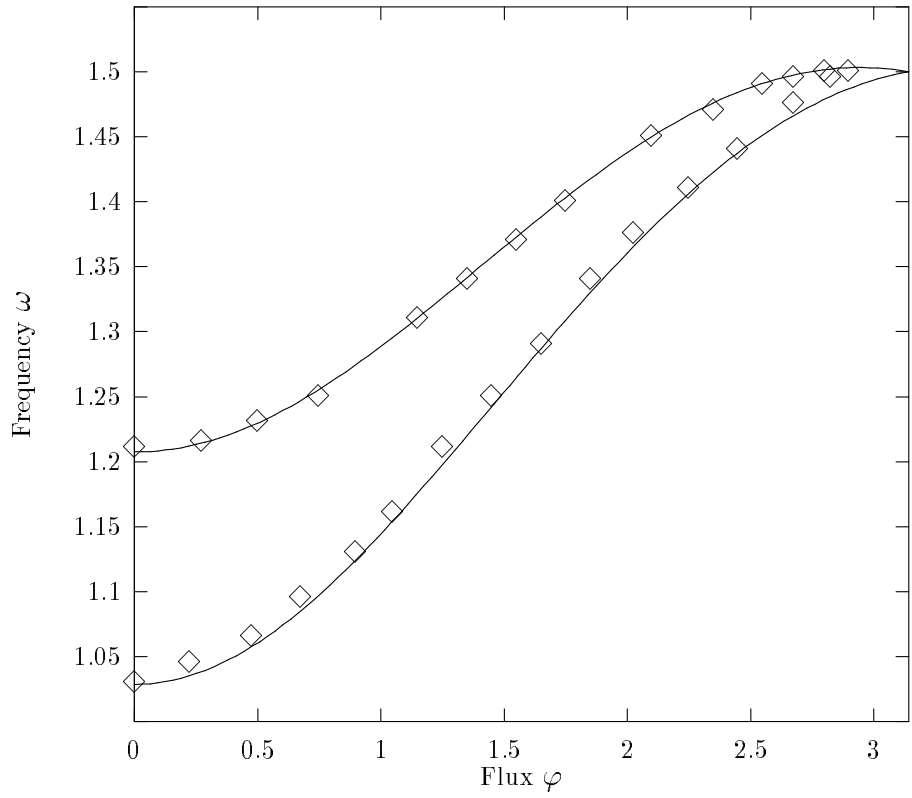
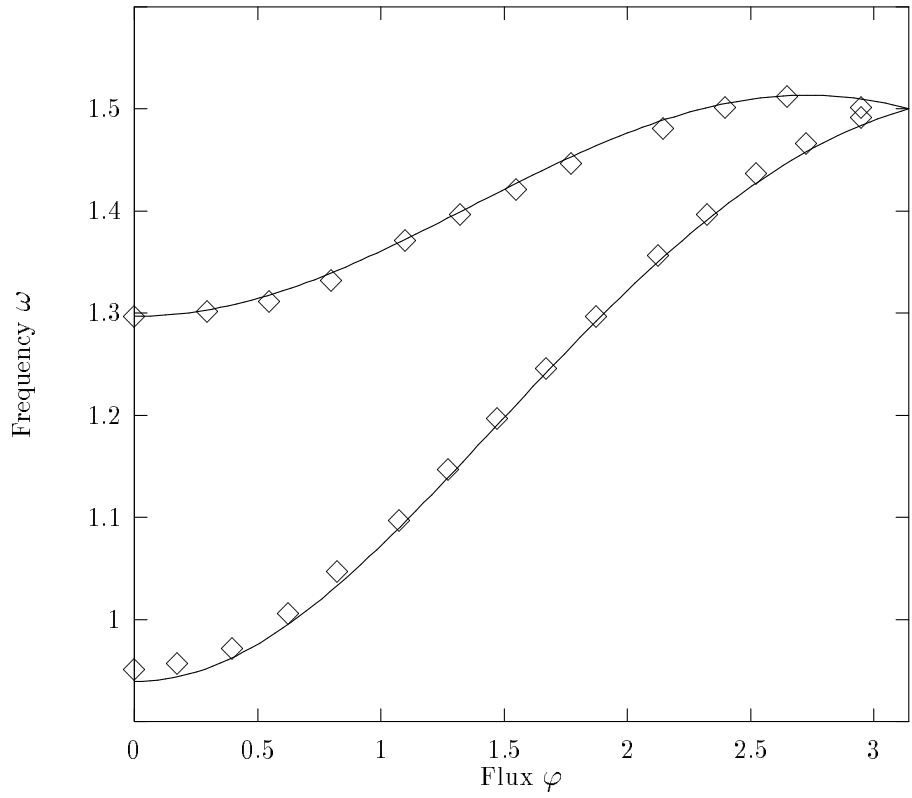
\end{document}